\documentclass[pre,twocolumn,english,aps]{revtex4-2}
\usepackage[T1]{fontenc}
\usepackage{tikz}
\usepackage{color}
\usepackage{amsmath}
\usepackage{amssymb}
\usepackage[latin1]{inputenc}
\usepackage{graphicx}
\usepackage{bm}
\usepackage{epsfig}
\usepackage{xcolor}
\usepackage[hidelinks]{hyperref}

\definecolor{mynicegreen}{RGB}{102,182,102}
\newcommand{\avg}[1]{\left\langle {#1} \right\rangle}

\begin{document}

\title{Sedimentation profiles and phase stacking diagrams in polydisperse hard rounded rectangle fluids}

\author{Tobias Eckert}
\affiliation{German Aerospace Center (DLR), Earth Observation Center, Remote Sensing Technology Institute, Oberpfaffenhofen, 82234 We{\ss}ling, Germany}
\affiliation{Physikalisches Institut, Universit{\"a}t Bayreuth, D-95440 Bayreuth, Germany}
\author{Daniel de las Heras}
\email{delasheras.daniel@gmail.com}
\homepage{www.danieldelasheras.com}
\affiliation{Institut f{\"u}r Theoretische Physik, Universit{\"a}t T{\"u}bingen, D-72076 T{\"u}bingen, Germany}
\affiliation{Physikalisches Institut, Universit{\"a}t Bayreuth, D-95440 Bayreuth, Germany}
\author{Enrique Velasco}
\affiliation{Departamento de F\'{\i}sica Te\'orica de la Materia Condensada, Instituto de F\'{\i}sica de la Materia 
Condensada (IFIMAC) and Instituto de Ciencia de Materiales Nicol\'as Cabrera, Universidad Aut\'onoma de 
Madrid E-28049 Madrid, Spain}
\author{Yuri Mart\'{\i}nez-Rat\'on}
\affiliation{Universidad Carlos III de Madrid, Departamento de Matem\'aticas, Grupo Interdisciplinar de Sistemas Complejos (GISC), 28911 Legan\'es, Spain}

\date{\today}

\begin{abstract}

We analyze the sedimentation behavior of a polydisperse two-dimensional liquid-crystal fluid using a local density functional theory based on scaled particle theory.
Polydispersity is incorporated through variations in the roundness of hard rectangular particles interacting solely via excluded area effects.
Despite its simplicity, the model displays a rich phenomenology.
In bulk, the fluid exhibits isotropic, nematic, and tetratic phases.
In sedimentation, we obtain complex phase stacking diagrams featuring multiphasic stacking sequences with up to four stacks of different bulk phases, inverted stacking sequences such as top isotropic and bottom nematic together with top nematic and bottom isotopic, as well as stacking sequences with reentrant stacks such as tetratic and nematic stacks floating between two isotropic stacks.
This phenomenology arises as a result of an intricate coupling between particle polydispersity and the effect of gravity.
Our approach can be easily adapted to investigate the sedimentation behaviour of other polydisperse colloidal systems.
\end{abstract}

\maketitle

\section{Introduction}

Colloidal systems are inherently polydisperse and exhibit variations in particle size and shape.
Polydispersity is more pronounced in some natural colloids, such as clays~\cite{Lagaly2006}. Recent improvements in synthetic methods have made it possible the synthesis of micronsize particles with sharp size distributions~\cite{Murphy2017,Kim2018,Roller2020,Voggenreiter2020}. 
However, a certain degree of polydispersity is unavoidable, even for microspheres~\cite{DeLaVega2013, Khlebtsov2023}, and small variations in particle sizes and shapes can affect the macroscopic properties of colloidal systems~\cite{Lekkerkerker2013,Almohammadi2025}. 

Polydispersity significantly alters the entropy, and consequently the bulk phase behavior of colloidal systems, which is often the result of a delicate balance between different entropic contributions to the free energy. In particular, the ideal mixing entropy increases with polydispersity, while the interaction (e.g. excluded-volume) contribution is modified in non-trivial ways by the presence of different species. Polydispersity changes the relative stability of bulk phases~\cite{Sollich2001,MartinezRaton2002,Filippo2023, Yuri2012, Velasco2014, Armas2017}, induces fractionation~\cite{Byelov2010}, and also preempts the formation of certain bulk phases that emerge in the corresponding monodisperse systems.
Above a terminal polydispersity, the crystallization of hard-spheres~\cite{Pusey1987,PhysRevE.59.618,Auer2001} and the formation of smectic phases in suspensions of colloidal rods~\cite{Bates1998} are absent. 
In suspensions of silica rods, polydispersity can suppress the formation of crystalline phases, favoring instead a smectic B phase~\cite{Kuijk2012}. 
Conversely, polydispersity can also stabilize phases that are not stable in the corresponding monodisperse system~\cite{Petukhov2005,Sun2009,Kotni2022}.
A detailed understanding of how polydispersity modifies the bulk behavior remains an ongoing challenge.

Valuable information about the bulk phase equilibria of colloidal suspensions can be obtained from sedimentation experiments, where a colloidal sample is left to equilibrate in a cuvette under the influence of gravity. However, the gravitational length (i.e., the ratio between thermal energy and gravitational energy per unit of height) if often comparable to or even smaller than the sample height in colloidal systems. 
Hence, there is frequently a strong coupling between the gravitational field and bulk phenomena due to the gravity-induced particle density gradient along the vertical direction. 
In monodisperse colloidal systems, the height dependent density distribution provides a direct and rather intuitive way to understand the bulk phase behavior~\cite{Biben1993, Piazza1993, Beek2004, Savenko2004, Eckert2023}.
In binary mixtures, there exist in general two distinct gravitational lengths and hence gravity affects each species differently.
The gravitational field stabilizes the formation of stacking sequences with several stacks of different bulk phases~\cite{Kooij1999,Kooij2000,Luan2009,Nakato2014,Chen2015, Eckert2021}. 
Moreover, the same bulk phase can appear twice within the cuvette.
An example is the isotropic-nematic-isotropic stacking sequence experimentally observed in sphere-plate mixtures~\cite{Heras2012}. 
Hence, drawing conclusions about bulk phenomena in binary mixtures from sedimentation experiments is a delicate issue.

An even more intricate interplay between gravity and bulk phenomena is expected in polydisperse colloidal systems since particle sizes and buoyant masses are not limited to a discrete set but are instead described by a density distribution function. 
The gravitational field creates a height-dependent density distribution function that differs from its bulk (parent) counterpart.
Hence, phases that are not stable in bulk for a given parent distribution might appear in a sedimentation sample due to e.g. gravity-induced strong fractionation. 
Sedimented samples of highly polydisperse goethite nanorods revealed the formation of a smectic phase~\cite{Vroege2006,vandenPol2008} even though the polydispersity of the parent distribution was well above the theoretical terminal polydispersity for smectic phases~\cite{Bates1998}.
A polydisperse suspension of natural clay rods developed nematic-nematic demixing due to a pronounced, gravity-induced, fractionation in the rod length~\cite{Zhang2006}.
In a suspension of highly polydisperse gibbsite platelets, the sedimented samples exhibited either an isotropic stack on top of a nematic one, or the inverse sequence (top nematic and bottom isotropic) depending on the sample height and average packing fraction~\cite{VanderKooij2001}.

To infer bulk phase equilibria from sedimentation experiments in polydisperse systems, we must first understand the role played by the gravitational field. 
From a theoretical point of view, sedimentation path theory~\cite{Heras2013,Geigenfeind2016,Eckert2021} incorporates the gravitational field on top of the bulk description of the system. 
The theory relies on a local equilibrium approximation that at each height maps the sedimented sample to a bulk system using local (height-dependent) chemical potentials.
So far, sedimentation path theory was used to study sedimentation in binary colloidal mixtures~\cite{Heras2012,Heras2013,Drwenski2016,PhysRevE.93.030601,Geigenfeind2016,Avvisati2017,Dasgupta2018,BrazTeixeira2021,Eckert2021,Eckert2022}, polymer-colloid mixtures~\cite{Heras2015}, and mass-polydisperse colloidal systems~\cite{Eckert2022b,Eckert2023}.
In a mass-polydisperse system, the colloidal particles have identical shapes and sizes but there is a continuous distribution of buoyant masses.
Since mass-polydispersity does not alter the interparticle interactions with respect to a monodisperse system, it is possible to derive an effective chemical potential that describes how the state of the mass-polydisperse sample changes along the cuvette. 
Mass-polydispersity does not affect bulk equilibria but it can have a profound impact on sedimented samples with mass-distributions close to the experimentally relevant density-matching regime~\cite{Eckert2022b, Eckert2023}. 

Studying mass-polydisperse models is useful to isolate the effect of the gravitational field but it does not allow us to understand the interplay between bulk equilibria and gravity.
As a first step in this direction, we theoretically investigate here the sedimentation of a two-dimensional model colloidal system with shape polydispersity by minimizing a local density functional in the presence of gravity.
Bellier-Castella and Xu used a conceptually similar approach to study sedimentation of polydisperse isotropic particles (Van der Waals fluids)~\cite{BellierCastella2003}.
We model the particles as polydisperse hard rounded rectangles (HRR).
Mart\'{\i}nez-Rat\'on and Velasco recently investigated the bulk  system~\cite{Yuri2022} with scaled particle theory (SPT)~\cite{Helfand1960}.
The model exhibits isotropic, tetratic, and nematic phases.
Several effects were attributed to polydispersity~\cite{Yuri2022}: (i) a decrease of the stability of the tetratic phase, (ii) the occurrence of strong fractionation between coexisting phases, (iii) a change in the nature of the isotropic-nematic bulk transition from continuous to first order, and (iv) a packing fraction inversion in which the disordered isotropic phase has higher packing fraction than the orientationally ordered nematic phase.

In this work, we focus on the effect of the gravitational field on a polydisperse fluid of HRR.
We first extend the scaled particle theory presented in Ref.~\cite{Yuri2022} to spatially inhomogeneous systems via a simple local density approximation.
Gravity is then incorporated to the polydisperse system as an external potential contribution to the free energy density functional. 
The height-dependent particle distribution function along the sedimented sample is obtained via a free minimization of the functional with respect to the full particle distribution resolved in both space and orientations.
We found several stacking sequences that we group in a stacking diagram in the plane of average packing fraction and sample height.
Unlike monodisperse and mass-polydisperse systems, shape polydisperse systems exhibit pairs of inverted stacking sequences, e.g.~tetratic-nematic and its inverse nematic-tetratic.
These inverted sequences are stable in different regions of the stacking diagram. 
Moreover, the degree of polydispersity has a strong effect on the topology of the stacking diagram, emphasizing its essential role on sedimentation.

\section{Model}

\begin{figure*}
    \includegraphics[width=0.9\linewidth]{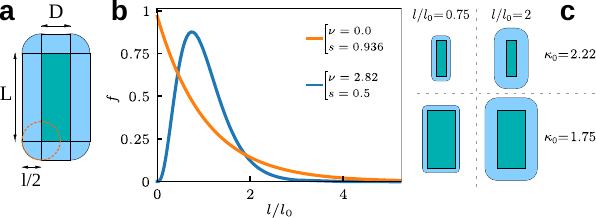}
	\caption{Particle model and distributions. (a) Sketch of a HRR (blue) obtained from a fixed rectangular core (green) of dimensions $L\times D$ by sliding a disk of diameter $l$ (dashed-orange) around the perimeter of the core. (b) Parent distributions, $f(l)$, as a function of the scaled roundness length $l/l_0$. The values of the parameters $\nu$ and $s$ characterizing the distributions are indicated in the figure. (c) Representative examples of the particles considered in this study: polydisperse distribution functions with mean aspect ratios $\kappa_0=2.22$ and $\kappa_0=1.75$, see Eq.~\eqref{eq:kappita}.
    }
	\label{fig1}
\end{figure*}

\noindent{\bf Particle model.} The fluid is composed of two-dimensional hard particles, each one with a fixed rectangular core of length $L$ and width $D$ 
($L\geq D$) and a hard envelope obtained by sliding the center of mass of a disk of diameter $l$ along the perimeter of the core, resulting 
in a rectangular shape with rounded corners.
A sketch of the particle is shown in Fig.~\ref{fig1}(a). 
The roundness length $l$ is treated as a polydisperse variable and therefore distributed according to a given probability distribution function.
For a given $l$, the particle area is 
\begin{align}
	a(l)=LD+\left(L+D\right)l+\frac{\pi}{4}l^2.
	\label{area}
\end{align}

\noindent{\bf Particle distribution.} We consider a column of height $H$ of an equilibrium fluid of polydisperse HRR sedimented according to the law of gravity.
The main magnitude characterizing the fluid behavior is the density profile, $\rho(l,z,\phi)$, of particles with roundness length $l$, located at (vertical) position $z$ (measured from the bottom of the sample), and with principal axes (parallel to the core side of length $L$) forming an angle $\phi$ with respect to the $z$-axis (the direction of the nematic or one of the 4-atic directors) of a fixed reference frame. 

Let $\rho_0$ denote the mean density of the whole sample. Then, the following constraint on the density profile holds:
\begin{align}
	\frac{1}{H}\int_0^H dz \int_0^{\pi} d\phi \rho(l,z,\phi)=\rho_0 f(l). \label{constraint}
\end{align}
Due to the head-tail symmetry of the particles, the angular integration can be restricted to the 
interval $[0,\pi]$. 
The function $f(l)$ is the fixed parent roundness length-distribution function, which we take as the truncated 
Schulz distribution
\begin{align}
	f(l)={\cal C}\left(\frac{l}{l_0}\right)^{\nu}e^{-\alpha l/l_0}\Theta\left(l_{\rm max} -l\right),\label{eq:parent}
\end{align}
where $l_{\rm max}$ is the cut-off for the maximum roundness length and $\Theta(x)$ is the Heaviside function.
The normalization constant ${\cal C}$ and the parameter $\alpha$, which depend on the exponent $\nu$ and $l_{\rm max}$, are calculated by imposing the normalization
\begin{align}
	\int_0^{l_{\rm max}} dl f(l)=1,
\end{align}
and fixing the value of the mean roundness length  
\begin{align}
\int_0^{l_{\rm max}} dl l f(l)=l_0.
\end{align}
We use the mean roundness length, $l_0$, as our unit of length.
The exponent $\nu$ together with the parameter $\alpha$ in Eq.~\eqref{eq:parent} control the degree of polydispersity which we quantify through the relative standard deviation  
\begin{align}
	s&\equiv \sqrt{\frac{\langle l^2\rangle_f}{l_0^2}-1},\label{eq:deviation}
\end{align}
with
\begin{align}
    \langle l^2\rangle_f &\equiv \int_0^{l_{\rm max}} 
	dl l^2 f(l). 
\end{align}

In our numerical calculations, all the integrals in $l$ were performed using a Gauss-Legendre quadrature with $101$ points for $\nu=0$ and $l_{\rm max}=5l_0$ 
resulting in the polydisperse coefficient $s=0.936$, and 
$81$ points for $\nu=2.82$ and $l_{\rm max}=3l_0$ resulting in $s=0.5$. These are the two cases selected
for analysis, corresponding to large and moderate polydispersities.
The parent distribution functions and illustrative particle shapes corresponding to them are shown in Fig.~\ref{fig1}(b) and Fig.~\ref{fig1}(c), respectively.

\section{Theory}
We use the SPT to approximate the interaction or excess part of the Helmholtz free energy per unit of area, which is supplemented by a local density approximation (the implications of this approximation are discussed later).
The excess-free  energy density (i.e., the excess part of the Helmholtz free-energy,  ${\cal F}_{\rm exc}$, divided by the system area, $A$, and scaled with the Boltzmann factor, $\beta^{-1}=K_B T$) depends locally on $\rho(l,z,\phi)$ and has the form \cite{Yuri2022}:
\begin{align}
	\Phi_{\rm exc}(z)&\equiv \frac{\beta {\cal F}_{\rm exc}[\rho]}{A}\nonumber\\
    &=-m_0^{(0)}(z) \ln\left[1-\eta(z)\right]+\frac{\langle\langle A_{\rm spt}\rangle\rangle(z)}
	{1-\eta(z)},\label{eq:Phiexc}
\end{align}
where $m_0^{(0)}(z)$ is the integrated density profile (see below), and the local packing fraction is defined as 
\begin{align}
	\eta(z)=\int_0^{l_{\rm max}} dl \int_0^{\pi} d\phi \rho(l,z,\phi) a(l).
	\label{eta}
\end{align}
Inserting Eq.~\eqref{area} into~\eqref{eta} we obtain 
\begin{align}
	\eta(z)=LD m_0^{(0)}(z) +(L+D)m_1^{(0)}(z)+\frac{\pi}{4} m_2^{(0)}(z),
\end{align}
where we have introduced the generalized Fourier moment profiles 
\begin{align}
	m_i^{(k)}\!(z)\!=\!\frac{2}{1+\delta_{k0}} \!
	\int_0^{l_{\rm max}}\! dl l^i \! \int_0^{\pi} \! d\phi \cos(2k\phi) \rho(l,z,\phi)\label{coefficients}
\end{align}
with
\begin{align}
k = \left\{
\begin{array}{ll}
	0 & \text{if } i =\{1,2\} \\
	\{0,1,\dots,n_{\rm max}\} & \text{if } i=0.
\end{array}
\right.
\end{align}
Here, $\delta_{kj}$ is the Kronecker delta.
The value $n_{\rm max}$ is chosen to ensure an adequate approximation for the orientational distribution function $h(z,\phi)$ (see below). 
Note that we only need the moments $\{m_i^{(0)}\}$ to define $\eta(z)$.
However, at this point, we introduce all the moments required to find the equilibrium density profile, $\rho(l,z,\phi)$, as will be shown later.

The magnitude $\langle\langle A_{\rm spt}\rangle\rangle$ in Eq.~\eqref{eq:Phiexc} is the double angular average, with respect to the density profiles $\rho(l,z,\phi)$ and $\rho(l',z,\phi')$, of the so called SPT-area integrated with respect to the polydisperse roundness lengths $l$ and $l'$:
\begin{align}
	\langle\langle A_{\rm spt}\rangle\rangle(z)\equiv\!\!&
	\int_0^{l_{\rm max}}\!\!\!dl\!\int_0^{l_{\rm max}}\!\!\!dl'\!\!\int_0^{\pi}\!\!\!d\phi\!\!\int_0^{\pi}\!\!\!d\phi' \rho(l,z,\phi)\nonumber\\
	&\times\rho(l',z,\phi')A_{\rm spt}(l,l',\phi-\phi'). 
	\label{double}
\end{align}
In turn, $A_{\rm spt}(l,l',\phi)$ can be calculated from the excluded area, $A_{\rm excl}(l,l',\phi)$, as
\begin{align}
	A_{\rm spt}(l,l',\phi)&=\frac{1}{2}\left[A_{\rm excl}(l,l',\phi)-a(l)-a(l')\right] \nonumber\\
	&=\frac{\left(L^2+D^2\right)}{2}|\sin\phi| 
	+LD|\cos\phi|\nonumber\\
	&+\frac{\left(L+D\right)}{2}(l+l')+
	\frac{\pi}{4}ll'. 
	\label{spta}
\end{align}
The excluded area, $A_{\rm excl}(l,l',\phi)$, between two HRRs with roundness lengths $l$ and $l'$ and orientation $\phi$ is the region of space inaccessible to the center of mass of one particle due to the presence of the other particle. Note that here $\phi$ is the relative orientation between both particles.

The Fourier expansion of the SPT-area~\eqref{spta} using the cosine basis functions $\{\cos(2k\phi)\}_{k=0}^{n_{\rm max}}$ is 
\begin{align} 
	A_{\rm spt}(l,l',\phi)&=\frac{\left(L+D\right)}{2}
	(l+l')+\frac{\pi}{4}ll'\nonumber\\
	&+\frac{1}{\pi}\left[g_0+2\sum_{k=1}^{n_{\rm max}}g_k\cos(2k\phi)\right],
	\label{Fourier}
\end{align}
with Fourier coefficients,
\begin{align}
    g_k&=-\frac{\left(L+(-1)^kD\right)^2}{4k^2-1}, \quad k=0,\dots,n_{\rm max}.
\end{align}
Inserting Eq.~\eqref{Fourier} into~\eqref{double} we arrive at
\begin{align}
	&\langle\langle A_{\rm spt}\rangle\rangle(z)=\frac{1}{\pi} \left[
		g_0\left(m_0^{(0)}(z)\right)^2+\frac{1}{2}\sum_{k=1}^{n_{\rm max}} g_k \left(m_0^{(k)}(z)\right)^2\right]\nonumber\\
	&+\left(L+D\right)m_0^{(0)}(z)m_1^{(0)}(z)+\frac{\pi}{4}
	\left(m_1^{(0)}(z)\right)^2,
\end{align}
where we have used the definitions of the generalized Fourier moment profiles~\eqref{coefficients}.

Equations~\eqref{eq:Phiexc} and~\eqref{double} constitute a local density functional approximation.
For this approximation to accurately take into account short ranged particle correlations, the variation of the integrated density profile, $m_0^{(0)}(z)$, along $z$ should be much less than the inverse of the average characteristic length of the interparticle potential, $\Lambda\equiv 2(L + l_0)$.
Hence, in our case, the relative variation of the density profile along $z$ should be of the order of the gravitational length $\langle\xi\rangle\equiv\left(\tau \langle a\rangle_f\right)^{-1}$. That is,
\begin{align}
	\frac{1}{m^{(0)}_0(z)}\left| \frac{dm_0^{(0)}(z)}{dz}\right|\alt \langle\xi\rangle^{-1}.
\end{align}
As $\langle\xi\rangle \gg\Lambda$, we obtain a condition to justify the local density approximation,
\begin{align}
	\frac{1}{m^{(0)}_0(z)}\left| \frac{dm_0^{(0)}(z)}{dz}\right|\ll \Lambda^{-1}.
\end{align}
We have defined 
\begin{align}
\langle a\rangle_f=LD+(L+D)l_0+\frac{\pi}{4} l_0^2(1+s^2)
\end{align}
as the averaged particle area with respect to the parent distribution function $f(l)$, while $\tau=\beta (\Delta d) g$ is a coefficient defined from the product  of $\Delta d\equiv d_{\rm m} -d_{\rm s}>0$, the difference between the mass density of the material from which solute particles are made ($d_{\rm m}$)  and that of the solvent ($d_{\rm s}$), and the constant of gravity $g$, divided  by $\beta^{-1}=k_B T$.

The ideal part of the free-energy density in reduced thermal units for the polydisperse mixture is given exactly by
\begin{align}
	\Phi_{\rm id}(z)&\equiv \frac{\beta{\cal F}_{\rm id}[\rho]}{A}\nonumber\\
    &=\int_0^{l_{\rm max}}\!\!dl\int_0^{\pi} d\phi 
	\rho(l,z,\phi)\left[\ln \left(\rho(l,z,\phi)\right)-1\right],
\end{align}
where we have dropped the (irrelevant) particle thermal areas.

Finally, the gravitational field is incorporated as an external potential contribution due to a conservative force field, 
\begin{align}
	\Phi_{\rm ext}(z)&\equiv \frac{\beta {\cal F}_{\rm ext}[\rho]}{A}\nonumber\\
    &=\tau z \int_0^{l_{\rm max}}\!\!dl\!\int_0^{\pi} d\phi 
	\rho(l,z,\phi) a(l)=\tau z \eta(z).
\end{align}

The total Helmholtz free-energy functional per unit of area in reduced thermal units is
\begin{align}
	\Phi[\rho]\equiv \frac{\beta {\cal F}[\rho]}{A}=\frac{1}{H}\!\int_0^{H}\!\!\!dz\left[
		\Phi_{\rm id}(z)+\Phi_{\rm exc}(z)+\Phi_{\rm ext}(z)\right].
	\label{total}
\end{align}

The functional minimization of $\Phi[\rho]$ with respect to $\rho(l,z,\phi)$, taking into account the constraint~\eqref{constraint}, yields
\begin{align}
	\rho(l,z,\phi)&=\frac{\rho_0 f(l) e^{c(l,z,\phi)}}
	{T(l)}, \label{result}\\
	c(l,z,\phi)&\equiv
	c_1(l,z,\phi)-\tau a(l)z,\\
	T(l)&\equiv H^{-1}\int_0^H dz'\int_0^{\pi}d\phi'e^{c(l,z',\phi')}
\end{align}
where the one-body direct correlation function is 
\begin{align}
	-c_1(l,z,\phi)&=-\ln\left[1-\eta(z)\right]+
	\frac{S(l,z,\phi)}{1-\eta(z)}+ p^*(z)a(l),
\end{align}
with
\begin{align}
	S(l,z,\phi)&\equiv 2\!\!\int_0^{l_{\rm max}}\!\!dl'\!\!\int_0^{\pi}\!\! d\phi' 
	\rho(l',z,\phi') A_{\rm spt}(l,l',\phi-\phi'),
\end{align}
and
\begin{align}
	p^*(z)&\equiv \frac{m_0^{(0)}(z)}{1-\eta(z)}+
	\frac{\langle\langle A_{\rm spt}\rangle\rangle(z)}{(1-\eta(z))^2},
\end{align}
being the local pressure profile in reduced thermal units, i.e.,~$p^*(z)=\beta p(z)$.
The function $S(l,z,\phi)$ in turn can be expressed as a function of the 
generalized Fourier moment profiles $\{m_k^{(i)}(z)\}$ taking into account the Fourier expansion 
of $A_{\rm spt}(l,l',\phi)$, given in Eq.~\eqref{Fourier}, resulting in 
\begin{align}
	S(l,z,\phi)=\frac{2}{\pi}\left[g_0m_0^{(0)}(z)+\sum_{k=1}^{n_{\rm max}} g_k m_0^{(k)}(z)\cos(2k\phi)\right]\nonumber\\
	+(L+D)m_1^{(0)}(z)+\left[(L+D)m_0^{(0)}(z)+\frac{\pi}{2}m_1^{(0)}(z)\right]l.
\end{align}
Using the result~\eqref{result} and the 
definitions~\eqref{coefficients}, we obtain the following set of self-consistent nonlinear integral 
equations for the unknown functions $\{m_i^{(k)}(z)\}$: 
\begin{align}
	&m_i^{(k)}(z)=\frac{2\rho_0}{1+\delta_{k0}}
	\int_0^{l_{\rm max}} dl l^i\frac{f(l)}{T(l)} \int_0^{\pi} d\phi 
	\cos(2k\phi) e^{c(l,z,\phi)}. \label{system}
\end{align}

This system of nonlinear equations allows us to find the $n_{\rm max}+3$ unknown generalized moment profiles $\{m_i^{(k)}(z)\}$; 
three $m_i^{(0)}(z)$ corresponding to $i=0,1,2$ and a total number of $n_{\rm max}$ moment profiles $m_0^{(k)}(z)$ for $1\leq k\leq n_{\rm max}$.
Here, we set $n_{\rm max}=20$.

The system of equations is solved iteratively, after uniform discretization of the spatial coordinate, 
using Anderson's acceleration method~\cite{Anderson} with memory length $M=3$, 
by gradually increasing the mixing parameter from $10^{-8}$ to $1024\times 10^{-7}$. 
The iterations stop when the residual (quadratic mean) is less than $10^{-6}$. To perform 
the integrals over $\phi$ we use the Gauss-Legendre quadrature with either $81$ or $101$ sample points in the interval 
$[0,\pi]$.
We perform the integration over $z$ using the Simpson's rule with $n_p=\max(\lceil 450\tau \rceil + 1, 41)$ uniformly distributed points, where $\lceil\cdot\rceil$ denotes the ceiling function.

The procedure to obtain the equilibrium density profile of a given sample is as follows.
First, we fix the average packing fraction, the sample height, and the parent distribution function.
Next, we initialize the generalized moment functions $\{m_i^{(k)}(z)\}$ with certain guesses depending on the bulk phases that we want to include along the column as possible candidates to the equilibrium profiles (see Sec.~\ref{parameters}).
Then, we iteratively solve the system~\eqref{system} to find the final equilibrium profile $\rho_{\rm eq}(l,z,\phi)$, which is a function of the set of equilibrium generalized moment functions.
In some cases different initial guesses result in different converged profiles.
To distinguish which of them is the equilibrium one, we compute the free energy for all converged profiles and choose the one with the lowest free energy. 
It is straightforward to show from Eqs.~\eqref{total},~\eqref{result}, and~\eqref{system} that the free energy at equilibrium is given by
\begin{align}
	\Phi[\rho_{\rm eq}]&=\rho_0\left(\ln \rho_0+\int_0^{l_{\rm max}} dl f(l) 
	\ln\left[\frac{f(l)}{T_{\rm eq}(l)}\right]\right)\nonumber\\
	&-\frac{1}{H}\int_0^H dz p^*_{\rm eq}(z).
	\label{equilibrium}
\end{align}

To end this section, we briefly remind the concepts of cloud and shadow coexisting phases at bulk.
These concepts will be used later when we compare the equilibrium sedimented phases obtained from the present model with those obtained at bulk conditions.
With cloud-A--shadow-B coexistence we mean that the whole sample filled by phase A (the cloud phase) coexists with an infinitesimally thin layer of phase B (the shadow phase).

\section{Sample characterization}
\label{parameters}

To describe the particle shape, we use both the mean aspect ratio, $\kappa_0$, and the roundness parameter, $\theta$, defined as 
\begin{align}
	\kappa_0&=\frac{L+l_0}{D+l_0},\label{eq:kappita}\\
    \theta&=\frac{l_0}{D+l_0}.\label{eq:theta}
\end{align}
The roundness parameter lies in the interval $[0,1]$ with $\theta\rightarrow0$ being the limit of a rectangle ($l_0\ll D$), and $\theta\rightarrow1$ being the limit of a discorectangle ($l_0\gg D$). 

Further, we define the scaled mean density $\overline{\eta}\equiv \rho_0\langle a\rangle_f$ or mean packing fraction (with respect to the parent distribution function) and 
the dimensionless density profile $\rho^*(z)\equiv m_0^{(0)}(z)\langle a\rangle_f$. 

To measure the degree of fractionation in the polydisperse roundness as a function of the elevation $z$, we define the distribution function $x(l,z)$ as the fraction of particles with roundness $l$ located in an infinitesimally thin slab at position $z$:
\begin{align}
	x(l,z)&\equiv \frac{\int_0^{\pi}d\phi \rho(l,z,\phi)}{\int_0^{l_{\rm max}} dl'\int_0^{\pi} 
	d\phi' \rho(l',z,\phi')}\nonumber\\
	&=\frac{1}{m_0^{(0)}(z)}\int_0^{\pi}d\phi \rho(l,z,\phi).
\end{align}
The distribution function $x(l,z)$ is normalized for all positions,
\begin{align}
	\int_0^{l_{\rm max}} dl x(l,z)=1,\ \forall \ z.
\end{align}
Then, we can compute the scaled average roundness (in units of $l_0$),
\begin{align}
\sigma_1(z)\equiv \langle l\rangle_x(z)/l_0,
\end{align}
and its mean square value 
\begin{align}
\sigma_2(z)\equiv \langle l^2\rangle_x(z)/l_0^2, 
\end{align}
as a function of the vertical coordinate $z$.
We define the moments of $l$ with respect to the distribution function $x(l,z)$ as 
\begin{align}
	\langle l^i \rangle_x(z)\equiv \int_0^{l_{\max}} dl l^i x(l,z)=\frac{m_i^{(0)}(z)}{m_0^{(0)}(z)},\quad i=1,2,
\end{align}
with the last equality obtained from~\eqref{coefficients}.
Both $\sigma_1$ and $\sigma_2$ help to characterize the mean roundness of particles populating the fluid slab at elevation $z$ together with the degree of fluid polydispersity at this position.   

We also use the maximum value of the distribution function $x(l,z)$ as a function of $l$ for a fixed $z$, ${\cal M}(z)\equiv \max_l \left[x(l,z)\right]$, to quantify the shape of $x(l,z)$ at each position $z$.

To find the stacking sequences~\cite{Heras2013} of the liquid crystal fluid column, it is necessary to describe the orientational symmetries of the different sedimented phases.
First, we define the orientational distribution function of the polydisperse fluid at the position $z$ as
\begin{align}
	h(z,\phi)\equiv \frac{\int_0^{l_{\rm max}} dl \rho(l,z,\phi)}{\int_0^{\pi} d\phi'\int_0^{l_{\rm max}}
	dl' \rho(l',z,\phi')},
\end{align}
which is normalized for all positions:
\begin{align}
\int_0^{\pi} d\phi h(z,\phi)=1,\ \forall \ z. 
\end{align}

Next, we define the orientational order parameters, measuring the degree of nematic ($n=2$) or tetratic ($n=4$) order, as cosines-weighted angular moments of the orientational distribution function
\begin{align}
	Q_{2n}(z)=\int_0^{\pi} d\phi h(z,\phi) \cos(2n\phi)=\frac{m_0^{(n)}(z)}{2m_0^{(0)}(z)},\quad n=1,2.
\end{align}
With the orientational order parameters, we classify the state of the sedimented sample at position $z$ according to the following rule:
\begin{itemize}
	\item Isotropic phase: $Q_2(z)=Q_4(z)=0$,
	\item Tetratic phase: $Q_2(z)=0$, $Q_4(z)> 0$.
	\item Nematic phase: $Q_2(z)>0$, $Q_4(z)> 0$.
\end{itemize}

\section{Results}

\subsection{Sedimentation profiles}

We begin by analyzing three sedimentation profiles, each corresponding to a distinct set of model parameters selected to illustrate representative behaviours of the system. Across all three cases, the following parameters are held constant: mean aspect ratio~\eqref{eq:kappita} $\kappa_0 = 1.75$, polydispersity coefficient~\eqref{eq:deviation} $s = 0.936$ (corresponding to $\nu=0$ with 
the parent distribution function shown in Fig.~\ref{fig1}), and mean roundness~\eqref{eq:theta} $\theta=0.3$. 
To generate the different profiles, we vary the capillary height $H$ and the mean packing fraction $\overline{\eta}$, allowing us to explore the influence of these parameters on the sedimentation behaviour.

\begin{figure*}
  \centering
  \includegraphics[width=0.9\linewidth]{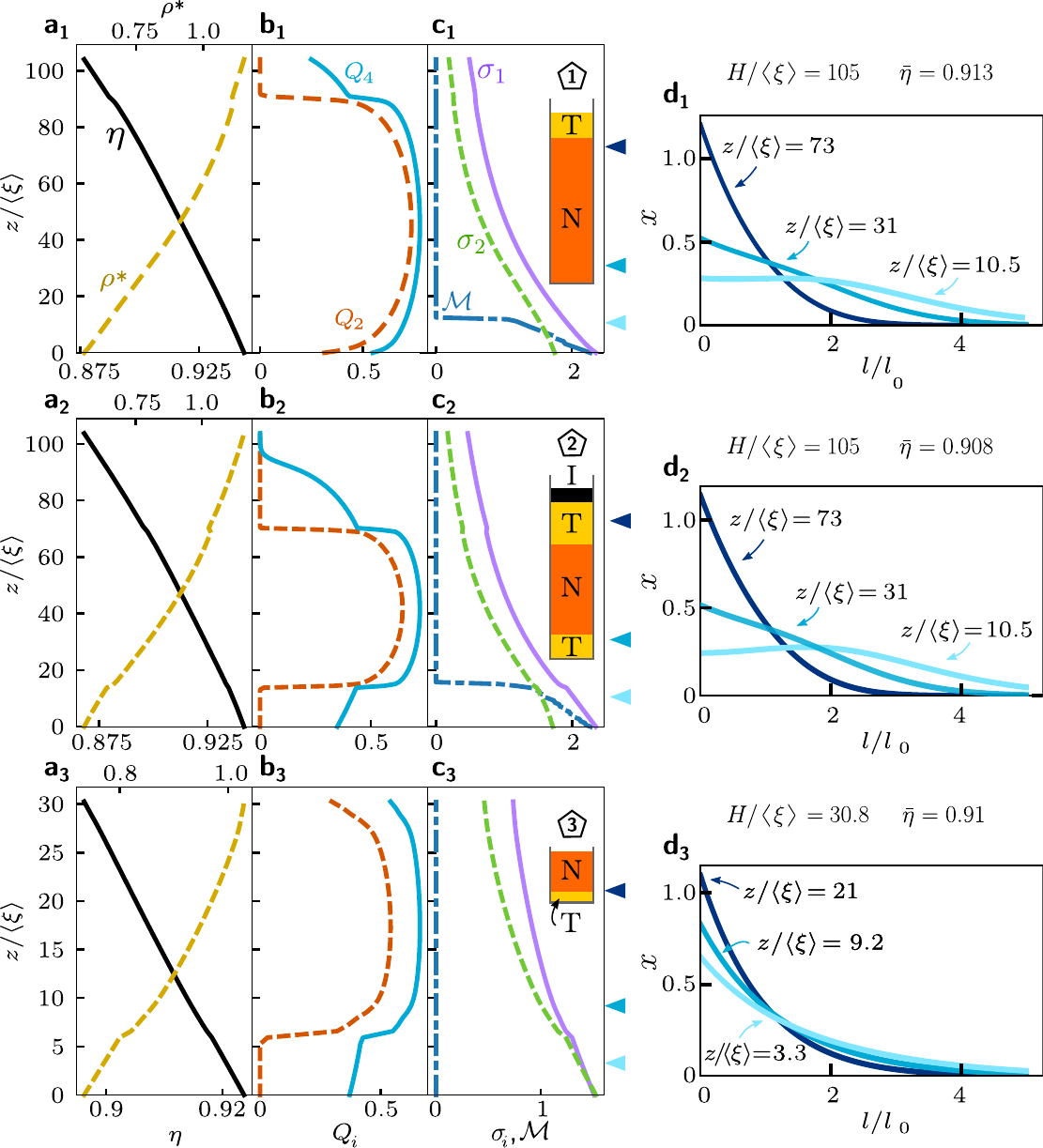}
  \caption{Sedimentation profiles. (a1) Scaled density $\rho^*$ and local packing fraction $\eta$ as a function of elevation $z$ for a sample with mean packing fraction $\overline{\eta}=0.913$ and sample height $H/\left<\xi\right>=105$, corresponding to a TN stacking sequence.
  The mean aspect ratio is $\kappa_0=1.75$, the polydispersity coefficient is $s=0.936$, and the mean roundness is $\theta=0.3$.
  (b1) Order parameters $Q_2$ and $Q_4$ as a function of elevation for the same sample as in (a1).
  (c1) Profiles for the first and second dimensionless moments $\sigma_1$, $\sigma_2$, and maximum value ${\cal M}$ of $x(l,z)$ with respect to $l$ as a function of $z$, for the same sample as in (a1).
  A sketch of the sample highlighting the stacking sequence is shown in panel (c1).
  (d1) Local roundness distribution functions $x(l,z)$ as a function of $l$ at three selected elevations $z$, marked in panels (c1) with arrows.
  Panels (a2) to (d2) display the same quantities as panels (a1) to (d1) for a sample with height $H/\left<\xi\right>=105$ and mean packing fraction $\overline{\eta}=0.908$, which corresponds to an ITNT stacking sequence.
  Panels (a3) to (d3) display the same quantities as panels (a1) to (d1) for a sample with height $H/\left<\xi\right>=30.8$ and mean packing fraction $\overline{\eta}=0.91$, which corresponds to a NT stacking sequence.
  Figure~\ref{fig3}(d), shows the location of the three samples in the stacking diagram using pentagons labeled 1 to 3.
  }
  \label{fig2}
\end{figure*}

Figure~\ref{fig2} displays the profiles along the sedimentation direction of the capillary for the three selected cases.
The first case corresponds to a sample height $H/\left<\xi\right>=105$ and the highest scaled mean density, $\overline{\eta} = 0.913$.
In this case, the profiles of the orientational order parameters, Fig.~\ref{fig2}(b1), reveal that in the top region, $Q_2 = 0$ and $Q_4 \ne 0$, while in the bottom region both $Q_2 \ne 0$ and $Q_4 \ne 0$.
Hence, the stacking sequence is TN (tetratic-nematic), where we denote the stacks of different bulk phases from top to bottom.

This sequence is consistent with the behaviour expected in bulk systems with low polydispersity.
In the central region of the capillary, the stable phase is the uniaxial nematic (N), although the uniaxial order parameter $Q_2$ decreases towards both the top and bottom boundaries.
This trend can be understood by considering the density and packing fraction profiles shown in Fig.~\ref{fig2}(a1), which indicates that the particle number density increases towards the top, opposite to the trend of the packing fraction.
This apparent contradiction arises due to gravity-driven sedimentation: heavier particles tend to settle lower in the capillary.
These heavier particles typically have higher roundness values $l$, and the aspect ratio $\kappa(l)$ is a monotonically decreasing function of $l$:
\begin{align}
    \kappa(l) = \frac{L+l}{D+l},
\end{align}
which implies that particles with greater roundness possess lower aspect ratios and are in turn more massive as Eq. (\ref{area}) for the particle area shows.
Since lower aspect ratios are associated with reduced uniaxial nematic order $Q_2$, the decrease in $Q_2$ towards the bottom is consistent with the presence of rounder, less anisotropic particles.
In contrast, the local packing fraction at the top is sufficiently reduced so as to stabilize the less ordered liquid-crystal T phase.
Figure~\ref{fig2}(c1) shows the first, $\sigma_1$, and second, $\sigma_2$, moments of the polydispersity distribution.
Both quantities increase towards the bottom of the capillary as a result of gravity-induced fractionation.
Near the bottom, this pronounced fractionation causes the peak of the roundness distribution to shift from $l = 0$ to significantly larger values, see Fig.~\ref{fig2}(d1).
That is, gravity induces a qualitative change in the shape of the local distribution of particles.

The stacking sequence shown in Fig.~\ref{fig2} panels (a2) to (d2) corresponds to the same value of sample height, $H/\left<\xi\right>=105$, but the scaled mean density is decreased to $\overline{\eta}=0.908$.
This small decrease is however sufficient to stabilise an isotropic stack at the top and a tetratic stack at the bottom, leading to a four stack ITNT sequence.
The bottom tetratic stack appears again as a result of the accumulation of more rounded (heavier) particles toward the bottom of the sample. 

Finally, in Fig.~\ref{fig2} panels (a3) to (d3) we show that by reducing the capillary height to $H / \langle \xi \rangle = 30.8$ and for an average packing fraction $\overline{\eta}=0.91$, the nematic phase no longer has sufficient vertical space to establish an interface with either a tetratic or an isotropic phase at the top.
Instead, a nematic-tetratic (NT) stacking sequence emerges.
This is an inversion of the stacking sequence with respect to the TN sequence shown in Fig.~\ref{fig2} panels (a1) to (d1).
Similarly, the cloud-N--shadow-T and shadow-N--cloud-T coexist in bulk at approximately the same polydispersity~\cite{Yuri2022}. 
However, the inversion of the stacking sequence is much stronger under sedimentation conditions as compared to the bulk.
Other phase inversion phenomena are present in the system, as discussed below.
A conceptually similar inversion of the stacking sequence (between isotropic-nematic and nematic-isotropic) was experimentally observed by van der Kooij et al. in polydisperse platelets~\cite{VanderKooij2001}.

In contrast to the previous cases, the peak of the distribution function remains fixed at $l = 0$ throughout the whole vertical column, see Fig.~\ref{fig2}(d3).

\subsection{Phase stacking diagrams}
\begin{figure*}
  \centering
  \includegraphics[width=\linewidth]{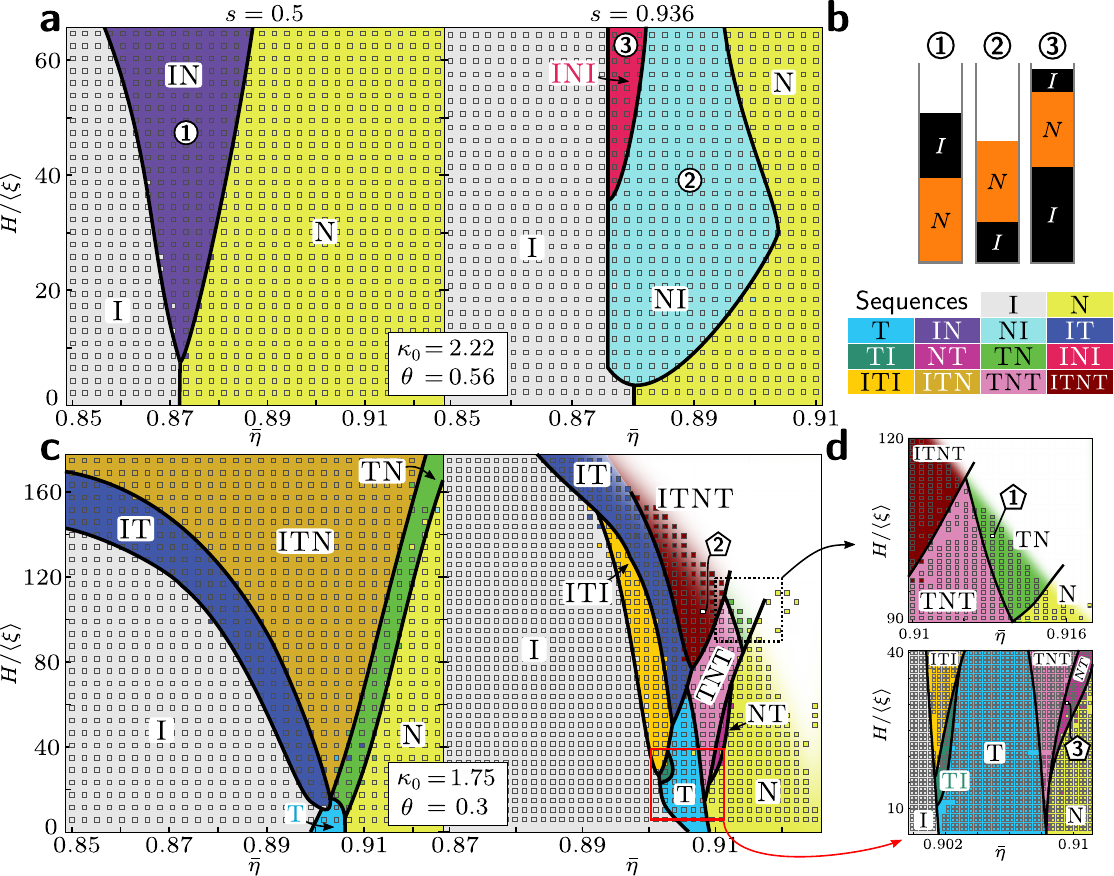}
  \caption{
    Stacking diagrams of polydisperse hard rounded rectangles in the plane of average packing fraction $\overline{\eta}$ and scaled sample height $H/\avg{\xi}$.
    The aspect ratio and average roundness of the parent (bulk) distributions are: (a,b) $\kappa_0=2.22$, $\theta=0.56$ and (c,d) $\kappa_0=1.75$, $\theta=0.3$.
    Two degrees of polydispersity are considered in each case: $s=0.5$ (a,c) and $s=0.936$ (b,d), as indicated above panels (a,b).
    Each gray square corresponds to a sedimentation sample that we calculated to obtain the stacking diagrams. 
    The black solid lines mark the approximate boundaries between two different stacking sequences in the stacking diagrams. 
    Stacking sequences are labeled from top to bottom and colored differently (see color box). 
    Close-up views of two highlighted regions in panel (d) are shown in the two side panels.
    The numbered black circles in (a,b) mark the position of the three sedimentation samples sketched to the right of panel (b).
    The labeled pentagons in (d) and white squares mark the position in the stacking diagram of the three sedimentation samples depicted in Fig.~\ref{fig2}. 
    Minimization of the functional was not possible in the white region of panel (d) due to numerical instabilities. 
    }
  \label{fig3}
\end{figure*}

We next group all possible stacking sequences for a given parent distribution in a stacking diagram, see Fig.~\ref{fig3}.
We construct the stacking diagram by computing sedimentation profiles for each point on a rectangular grid in the plane defined by the average packing fraction $\overline{\eta}$ and the sample height $H$.
The phase of each horizontal stack is then identified using the orientational order parameters $Q_2(z)$ and $Q_4(z)$.
The grid size is chosen according to the complexity of the diagram.
A finer grid is used when the number and symmetries of the involved stacking sequences is rather sensitive to small changes in $\overline{\eta}$ and/or $H/\langle \xi\rangle$.
Each point in the stacking diagram corresponds to a sedimented sample with a given stacking sequence.
As mentioned above, we label the stacking sequences from top to bottom.
Hence, the stacking sequence INT means that the stacks I, N and T are observed from the top towards the bottom along the vertical direction.
Three schematic examples of biphasic (1 and 2) and triphasic (3) stacking sequences are represented next to panel (b) of Fig. \ref{fig3}.
These particular samples are indicated in the corresponding stacking diagrams shown in Fig.~\ref{fig3}(a) and Fig.~\ref{fig3}(b).

Four stacking diagrams are shown in Fig.~\ref{fig3} in the $\overline{\eta}$-$H/\avg{\xi}$ plane, where again $\avg{\xi} = (\tau\avg{a})^{-1}$ is the average gravitational length.
First, we study HRR systems with an average aspect ratio $\kappa_0 = 2.22$ (for which the tetratic phase T is not stable in bulk~\cite{Yuri2022}) under gravity, see Fig.~\ref{fig3}(a) and Fig.~\ref{fig3}(b).
For a moderate degree of polydispersity, $s=0.50$ (corresponding to the parent distribution function shown in Fig, \ref{fig1}), and $\theta=0.556$, we only find the stacking sequence IN (sketched sequence 1) in addition to the pure I and N stacks, see Fig.~\ref{fig3}(a).
For this parent distribution, the I-N transition is of first order in bulk and there is no phase inversion with respect to packing fraction~\cite{Yuri2022}.
Therefore, the stacking sequences found correspond to those expected for a bulk first-order transition.
For example, for $H/\langle \xi\rangle\approx50$ and mean packing fractions between $0.85$ and $0.9$, we observe the expected cascade of stacking sequences I $\to$ IN $\to$ N.
The region occupied by the sequence IN in the stacking diagram will presumably keep growing as we increase the sample height. 
In the limit of very large samples, the IN should develop at essentially any value of the average packing fraction.
The parent distribution contains only particles with positive buoyant masses and therefore in the limit $H/\langle \xi\rangle\rightarrow\infty$ the top stack must always be a dilute isotropic phase.
We expect that the stacking sequence IN is still present in the bulk limit, i.e.,~$H/\langle \xi\rangle \rightarrow 0$, with the corresponding density gap at coexistence.
For this parent distribution the most ordered bulk phase (N) appears at the bottom.
No inversion of the stacking sequence is present.

Using the same values for $\kappa_0$ and $\theta$, but increasing the polydispersity to $s\to 1$, we observe that no inversion in packing fraction occurs in bulk~\cite{Yuri2022}. 
Specifically, at the cloud-I--shadow-N coexistence, the packing fraction of the isotropic phase (I) is lower than that of the nematic phase (N).
For the cloud-N--shadow-I coexistence, both packing fractions are approximately equal.
However, the situation changes in presence of the gravitational field, see Fig.~\ref{fig3}(b).
When the polydispersity increases to $s = 0.936$, a phase inversion in the local packing fraction emerges: the conventional IN stacking sequence disappears (within the resolution of our sampling grid and also for the range of samples heights and packing fractions considered), and new sequences, NI and INI, are stabilized, see the sketched sequences 2 and 3 in Fig.~\ref{fig3}(b).
The sequence INI, i.e., a nematic stack floating between two isotropic stacks, has been experimentally observed in plate-rod binary colloidal mixtures~\cite{Heras2012} and theoretically studied with sedimentation path theory~\cite{Heras2012,Eckert2022}.

Although the NI sequence is present at all sample heights considered here, it will disappear above a certain height because in the limit $H/\langle \xi\rangle \rightarrow \infty$ a dilute isotropic must develop on top (note that all buoyant masses are positive).
The triphasic INI sequence occurs only for sample heights above $H / \langle \xi \rangle \gtrsim 35 $.
Hence, at a fixed sample height $H/\langle\xi\rangle = 60$ and increasing the mean packing fractions in the range $0.86 \le \overline{\eta} \le 0.9$, we observe the cascade I $\to$ INI $\to$ NI $\to$ N.

In the triphasic INI configuration, the more disordered I phase resides at the bottom, followed by an intermediate N layer, and then another I layer at the top, similar to the INI sequence found in sphere-plates binary mixtures~\cite{Heras2012,Eckert2022}.
There, the bottom (top) isotropic stack is rich in the heavier spheres (lighter plates).
Here, both isotropic stacks are occupied by HRRs that differ on their relative roundness size.
The INI sequence found here provides another direct evidence of a genuine phase inversion induced by the coupling between polydispersity and gravity, an effect that is absent in the bulk phase diagram for the same parameters~\cite{Yuri2022}.
Note also that no evidence of three-phase coexistence is present in bulk, highlighting the role of the gravitational field in the emergence of these complex stacking sequences.
The stabilization of the INI sequence requires a finite sample height ($H/\langle \xi \rangle \ne 0$) which clearly distinguishes this phenomenon from bulk behavior.

In summary, even when only two stable bulk phases, I and N, do exist, we observe a clear inversion in the stacking sequence (from IN to NI) and the emergence of a reentrant I stack within the INI sequence as the degree of polydispersity $s$ increases.

We next reduce the average aspect ratio to $\kappa_0 = 1.75$, with the goal of introducing a new stable phase in bulk: the tetratic phase (T).
Under these conditions, we analyze a system with three stable bulk phases, namely the isotropic (I), tetratic (T), and nematic (N), using a roundness parameter $\theta = 0.3$ and a moderate degree of polydispersity, $s = 0.50$.
In bulk, the I-T transition is of second order, while the T-N transition is of first order.
No packing fraction inversion phenomena are observed in bulk~\cite{Yuri2022}.

In the sedimented system, see Fig.~\ref{fig3}(c), alongside pure I, T, and N stacking sequences, a rich variety of biphasic stacking sequences, IT and TN, as well as a triphasic stacking, ITN, are stabilized.
However, no inversion of the stacking sequence is present.
The most ordered stack appears always at the bottom.
The transitions between stacking sequences can be quite intricate: for example, at $H/\langle \xi \rangle \approx 80$, we observe the cascade I $\to$ IT $\to$ ITN $\to$ TN $\to$ N as $\eta$ increases over a relatively narrow interval ($0.88 \lesssim \overline{\eta} \lesssim 0.93$).
Notably, the ITN stacking sequence disappears if the sample height is smaller than $H/\langle \xi \rangle \approx 20$, indicating that the corresponding I-T-N three-phase coexistence does not occur in bulk.
We anticipate that the region corresponding to the ITN sequence in the stacking diagram will continue to expand as the sample height increases.

For a higher degree of polydispersity, $s = 0.936$, an even wider variety of stacking sequences emerges, including IT, TI, ITI, TN, NT, TNT, and ITNT, see Fig.~\ref{fig3}(d).
It is interesting to note that inverted sequences such as IT and TI can appear at the same average packing fraction ($0.902 \le \overline{\eta} \le 0.904$) but for different sample heights.
Similarly, an inversion from TN to NT is also observed at $\overline{\eta} \sim 0.912$: the TN sequence appears in taller samples ($H/\langle \xi \rangle \sim 100$), while the NT sequence is found in shorter samples ($H/\langle \xi \rangle \sim 60$).
The samples shown in figure~\ref{fig2} panels (a1-c1) and (a3-c3) are illustrative examples of this inversion.
Two inverted sequences do not share a common boundary in the stacking diagram.
For example, transforming the sequence TN into NT (e.g., by increasing the sample height at constant packing fraction) requires passing through the intermediate region TNT in the stacking diagram. 

The topology of the phase stacking diagram is rather complex.
For example, if we fix $H/\langle\xi\rangle=100$ and vary $\overline{\eta}\in[0.89,0.92]$, then up to 7 stacking sequences appear as $\overline{\eta}$ increases: I $\to$ ITI $\to$ IT $\to$ ITNT $\to$ TNT $\to$ TN $\to$ N, with one of them including up to four stacks of different bulk phases along the column.
Also, if we fix the average packing fraction at $\overline{\eta}\approx0.906$ and increase the sample height within the interval $H/\langle\xi\rangle\in[0,160]$ we find the stacking sequences I $\to$ T $\to$ TI $\to$ ITI $\to$ IT $\to$ ITNT.
These examples illustrate the challenges in inferring bulk phase equilibria from a collection of sedimented samples.

To better visualize the stacking sequences that appear in a narrow range of $\overline{\eta}$, we have included two insets in Fig.~\ref{fig3}(d).
The first inset focuses on relatively large values of the sample height (including the discussed case of $H/\langle\xi\rangle=100$), while the second inset considers short samples with $H$ values up to $40\langle\xi\rangle$.
In the latter case, setting $H/\langle\xi\rangle\approx30$, we again observe a cascade of up to seven different stacking sequences: I $\to$ ITI $\to$ TI $\to$ T $\to$ TNT $\to$ NT $\to$ N (when $\overline{\eta}$ varies between 0.9 and 0.915).
An inversion in the local packing fraction is observed, with a T stack (having less orientational order than phase N) occupying to the bottom (in the case of the TNT or the NT sequence), and also an I stack (without orientational order) appearing at lower height than the T stack in the ITI sequence, where the phase I also reappears at the top of the column as a reentrant stack.

Two three-phase sequences (ITI and TNT) appear under sedimentation conditions but are absent in bulk ($H / \langle\xi\rangle \rightarrow 0$).
As we previously described, the local inversion in the packing fraction occurs because the more disordered phases are richer in particles with smaller aspect ratios and consequently more rounded, which in turn implies a greater area and hence more gravitational attraction, making them sink to the bottom.
The reentrant top isotropic (tetratic) stack in the INI (TNT) sequence occurs due to reduced number of particles in the upper portion of the column.
As density decreases, orientational order declines because interparticle interactions become less prominent. 

In the large-sample limit, multiple scenarios are plausible. There, the stacking sequences must develop a top isotropic stack followed by tetratic and nematic stacks. 
However, due to the occurrence of local phase inversion, the sequences ITNT and even ITNTI (not observed in the range of heights considered here) could be present and possibly dominate the stacking diagram at large-sample heights.

\section{Conclusions}

We have investigated the sedimentation behavior of a polydisperse two-dimensional liquid-crystal model using a local density-functional theory. 
This is the first consistent treatment of the intricate interplay between polydispersity, phase equilibria, and gravity in liquid-crystal fluids. 
Given the inherent complexity of the problem, we have employed a simplified particle model, rounded hard rectangles, in our theoretical framework.
Nevertheless, the same theoretical framework can be used to study other types of interaction potentials.
We argue that this minimal model is sufficient to capture qualitatively correct features, particularly in high columns composed of phases that are uniform in bulk.
Note, however, that our local density approximation assumes that the walls do not affect the properties of the fluid inside the capillary. Surface effects such as wetting, anchoring, and layering will modify the free-energy landscape and enrich the phenomena observed here.
Also, situations where the local-density approximation may not be accurate include the occurrence of stacks with positionally ordered phases (see below), and stacking sequences where the thickness of the stacks is of the same order of magnitude as the thickness of the interfaces that separate the corresponding stacks.
In these cases a nonlocal version of the theory would be necessary to quantitatively describe the sedimentation profiles.
In the stacking diagram, such sequences with narrow stacks occur near the boundaries between two stacking sequences.
However, we expect that the local approximation will qualitatively capture the stacking sequences and the topology of the stacking diagram. 

Despite the simplicity of the model, our analysis reveals a remarkably rich variety of stacking sequences, far exceeding the structural diversity observed in bulk. These include inverted sequences, reentrant stacks, multiphasic stacking sequences with up to four stacks, and cascades of transitions between different stacking sequences occurring by either varying the packing fraction while keeping the sample height constant or varying the height at constant average packing fraction. All of these phenomena arise from the coupling between particle polydispersity and the gravitational field.

In the present study we have used exponentially decaying parent distribution functions.
In principle other distributions with an exponential or more rapidly decaying functional forms will produce qualitatively similar results. The case is different when the distribution has a fat tail such as in log-normal or power-like distributions. For instance, in length-polydisperse rods, the more ordered nematic phase is enriched in particles of very large size, making the order of the phase transitions and fractionation effects much stronger~\cite{Sollich2003}. For our HRR model, we expect fat tails to enrich the I and T phases with particles of very large roundness.

Our two-dimensional model should also accurately represent the sedimentation behavior in tilted monolayers~\cite{Thorneywork2017} of polydisperse colloidal particles. 
Also, new experiments on tilted vibrated monolayers of granular rods can be designed to study the effect of gravity on the sedimentation behavior of granular particles.
Moreover, our approach, based on projecting the infinite-dimensional thermodynamic space of the polydisperse system onto a finite set of density moments, is computationally efficient, and readily extendable to three-dimensional colloidal systems.
Hence, it is a promising tool for studying more realistic systems where the coupling between gravity and polydispersity is significant.
Comparison with already available experimental data~\cite{VanderKooij2001,Zhang2006,vandenPol2008} should then be possible.
Our theoretical framework has potential applications ranging from better understanding the sedimentation of natural suspensions such as clay, which is crucial for erosion prediction, to optimizing the fabrication of colloidal inks where controlled particle settling prevents clogging, and modeling the sedimentation of biofluids which could improve diagnostics accuracy.

An interesting line of research for future works is to study the effect of continuous particle length polydispersity on the stacking diagrams of colloidal suspensions with stable non-uniform bulk phases, such as smectic, columnar, and crystalline phases.
Non-local density functionals, such as those based on the fundamental measure theory~\cite{Roth2010,MartnezRatn2011,Wittmann2016,Wittmann2017,ElMoumane2024}, are then required to accurately describe the bulk.
A full minimization of the free energy of the inhomogeneous system, as we have done in this work, would fully incorporate the effect of the gravitational field.
The frozen or discrete~\cite{MartinezRaton2002,Zwanzig1963} orientation approximations can alleviate the high computational cost required for the numerical implementation of non-local density functionals.
Still, accurately describing non-homogeneous bulk phases requires a computational grid with sub-particle resolution.
Hence, the minimization of a polydisperse non-local density functional in presence of gravity can only be done for relatively small systems~\cite{BellierCastella2003,Buzzacchi2004,Yu2004}, with a maximum height of a few hundred particle sizes.
This limits the ability to compare with standard sedimentation experiments in which the sample height is often thousands and even millions of times larger than the particle size.

An extension of equilibrium~\cite{Sammller2023} and non-equilibrium~\cite{Zimmermann2024} neural functionals to polydisperse systems could help address this issue, and also open the door to accurately describe the dynamics of sedimentation in polydisperse fluids.
Another promising possibility to overcome this limitation is to extend sedimentation path theory~\cite{Heras2013, Geigenfeind2016, Eckert2021} from mass-polydisperse~\cite{Eckert2022b} to fully polydisperse systems.
A sedimented sample would then be discretised using a relatively small number of horizontal slabs. 
Each slab would be approximated by a bulk system with the same local particle distribution as that in the slab.
An iterative procedure, conceptually similar to that done for mass-polydisperse systems~\cite{Eckert2022b}, could be used to find the mapping between the bulk and the set of horizontal slabs via a height-dependent distribution of local chemical potentials.
Sedimentation path theory assumes that each slab is equivalent to an equilibrium system (in contrast to the approach presented here) but it allows the study of arbitrarily large samples.
In addition, this local equilibrium approximation is usually accurate in colloidal systems.
A further advantage of sedimentation path theory is that it incorporates the effect of the gravitational field using only the bulk equation of state, regardless of its origin.
Hence, it might be possible to accurately describe the bulk of the polydisperse system combining computer simulation data and deep learning~\cite{Lijie2025}.
Moreover, sedimentation path theory would straightforwardly describe sedimentation in systems with stable non-uniform bulk phases,
provided that the underlying theory used to describe the bulk of the polydisperse system correctly accounts for such phases.

\begin{acknowledgements}
EV and YMR acknowledge financial support from Grants PID2023-148633NB-I00/AEI and PID2021-126307NB-C21/MICIU/AEI/10.13039/501100011033/FEDER, UE respectively. DdlH acknowledges support through the Heisenberg program of the Deutsche Forschungsgemeinschaft (DFG) under project number 550390029.
\end{acknowledgements}

\end{document}